\documentclass[a4paper,fleqn,usenatbib]{mnras}
\usepackage{newtxtext,newtxmath}
\usepackage[T1]{fontenc}
\usepackage{ae,aecompl}
\usepackage{graphicx}
\usepackage{hyperref}

\newcommand{\MSUN}{\,\mbox{$\mathrm{M_{\odot}}$}}
\newcommand{\RSUN}{\,\mbox{$\mathrm{R_{\odot}}$}}
\newcommand{\Ion}[2]{#1{\,\sc#2}}

\title[M dwarfs in close binaries]{The scatter of the M dwarf mass-radius relationship}

\author[S. G. Parsons et al.]{S.~G.~Parsons$^{1}$\thanks{s.g.parsons@sheffield.ac.uk},
B.~T.~G{\"a}nsicke$^{2}$,
T.~R.~Marsh$^{2}$,
R.~P.~Ashley$^{2}$,
E.~Breedt$^{3}$,
\newauthor
M.~R.~Burleigh$^{4}$,
C.~M.~Copperwheat$^{5}$,
V.~S.~Dhillon$^{1,6}$,
M.~J.~Green$^{2}$,
\newauthor
J.~J.~Hermes$^{7}$\thanks{Hubble Fellow},
P.~Irawati$^{8}$,
P.~Kerry$^{1}$,
S.~P.~Littlefair$^{1}$,
A.~Rebassa-Mansergas$^{9,10}$,
\newauthor
D.~I.~Sahman$^{1}$,
M.~R.~Schreiber$^{11}$
and M.~Zorotovic$^{11}$
\\
$^{1}$ Department of Physics and Astronomy, University of Sheffield,
Sheffield, S3 7RH, UK\\
$^{2}$ Department of Physics, University of Warwick, Coventry CV4 7AL, UK\\
$^{3}$ Institute of Astronomy, University of Cambridge, Madingley Road,
Cambridge CB3 0HA, UK\\
$^{4}$ Department of Physics and Astronomy, University of Leicester, Leicester
LE1 7RH, UK\\
$^{5}$ Astrophysics Research Institute, Liverpool John Moores University, IC2,
Liverpool Science Park, L3 5RF, UK\\  
$^{6}$ Instituto de Astrof{\'i}sica de Canarias, V{\'i}a Lactea s/n, La
Laguna, E-38205 Tenerife, Spain\\
$^{7}$ Department of Physics and Astronomy, University of North Carolina,
Chapel Hill, NC 27599-3255, USA\\
$^{8}$ National Astronomical Research Institute of Thailand, 191 Siriphanich
Bldg., Huay Kaew Road, Chiang Mai 50200, Thailand\\
$^{9}$ Departament de F\'isica, Universitat Polit\`ecnica de Catalunya,
c/Esteve Terrades 5, 08860 Castelldefels, Spain\\
$^{10}$ Institut d'Estudis Espacials de Catalunya, Ed. Nexus-201, c/Gran
Capit\`a 2-4, 08034 Barcelona, Spain\\
$^{11}$ Instituto de F{\'i}sica y Astronom{\'i}a, Universidad de
Valpara{\'i}so, Avenida Gran Bretana 1111, Valpara{\'i}so, 2360102, Chile\\
}

\date{Accepted 2018 August 23. Received 2018 August 22; in original form 2018 July 19.}

\pubyear{2018}

\begin{document}
\label{firstpage}
\pagerange{\pageref{firstpage}--\pageref{lastpage}}
\maketitle

\begin{abstract}
M dwarfs are prime targets in the hunt for habitable worlds around other
stars. This is due to their abundance as well as their small radii and low
masses and temperatures, which facilitate the detection of temperate, rocky
planets in orbit around them. However, the fundamental properties of M dwarfs
are difficult to constrain, often limiting our ability to characterise the
planets they host. Here we test several theoretical relationships for M dwarfs
by measuring 23 high precision, model-independent masses and radii for
M dwarfs in binaries with white dwarfs. We find a large
scatter in the radii of these low-mass stars, with 25 per cent having radii
consistent with theoretical models while the rest are up to 12 per cent
over-inflated. This scatter is seen in both partially- and fully-convective
M dwarfs. No clear trend is seen between the over-inflation and age or 
metallicity, but there are indications that the radii of slowly rotating M
dwarfs are more consistent with predictions, albeit with a similar amount
  of scatter in the measurements compared to more rapidly rotating M
  dwarfs. The sample of M dwarfs in close binaries with white dwarfs appears
indistinguishable from other M dwarf samples, implying that common envelope
evolution has a negligible impact on their structure. We conclude that
theoretical and empirical mass-radius relationships lack the precision and
accuracy required to measure the fundamental parameters of M dwarfs well
enough to determine the internal structure and bulk composition of the planets
they host.
\end{abstract}

\begin{keywords}
stars: late-type -- binaries: eclipsing -- stars: fundamental parameters
\end{keywords}

\section{Introduction}

The discovery of a super-Earth orbiting the nearby (14.6\,pc) M4.5 dwarf
GJ\,1214 \citep{charbonneauetal09-1} via photometric follow-up of
individual M-dwarfs \citep{nutzman08} and the recent radial-velocity detection
of an Earth-mass planet at Proxima Centauri \citep{anglada-escude16-1} demonstrates the enormous potential of planet searches focusing on low-mass stars, as their small radii and low masses substantially facilitate the discovery of smaller planets compared to planet searches at FGK stars. 
Consequently, M-dwarfs are now key targets of many transit and radial surveys,
e.g. NGTS \citep{wheatley18}, SPECULOOS \citep{delrez18} and CARMENES \citep{reiners18}. In particular, TESS will survey the brightest and closest M-dwarfs for transiting planets \citep{rickeretal15-1}, substantially increasing the number of known exoplanets orbiting low-mass stars \citep{ballard18-1}. 
The identification of several temperate Earth-sized planets orbiting low-mass
stars \citep{dittmannetal17-1, gillonetal16-1, gillonetal17-1}, combined with
the fact that M-dwarfs are the most numerous stars in the Milky Way, has led
to considerable interest in the habitability of these worlds
\citep{seager13-1, wandel18-1, kopparapuetal17-1}. 

A fundamental limitation in the characterisation of exoplanets is that the
derived bulk parameters, including masses, radii, and densities, require
accurate knowledge of the planet host properties. Accurate planet radii and
masses (which require accurate stellar radii and masses) are required to gauge
insight into their internal structure and bulk composition. 
\citet{valenciaetal07-1} argued that planet radius measurements to better than
5\% and mass measurements to better than 10\% are necessary to distinguish
between rocky and icy bulk composition, and even then, details of the interior
composition are model-dependent \citep{rogers+seager10-1, dornetal15-1}.

It has been well established that the measured radii of low-mass stars
($<$0.6\,\MSUN) are larger than predicted by evolutionary models, by up to
10-20 per cent \citep{lopezmorales05}. This is thought to be caused by the
fact that virtually all precise mass-radius measurements of low-mass stars
come from stars in close binaries\footnote{While accurate parallaxes help to
  constrain stellar radii of single stars if their effective temperature can
  be empirically constrained, their masses require additional information,
  such as an independent measure of their surface gravity from planet transits
  \citep{stassunetal17-1, southworthetal07-6}, or from their
  granulation-driven variability \citep{stassunetal18-1} and hence remain
  limited in their accuracy.}.
These stars are tidally locked and are
hence rapid rotators and magnetically active. This activity is thought to
lead to a cooler and larger star \citep{morales08} and can therefore explain
the over-inflation, an idea supported by the fact that the
interferometrically-measured radii of isolated, inactive low-mass stars appear
more consistent with evolutionary models \citep{demory09}. Magnetic activity
can also explain the 14 per cent larger radii of young low-mass stars in the
Pleiades cluster \citep{jackson18}. However, the reality is more complicated
than this, as there are several relatively inactive nearby low-mass stars with
interferometric radii more than 15 per cent too large \citep{berger06} and
there are stars in long period, slowly rotating binaries that are also oversized
\citep{doyle11,irwin11}. Conversely, there are rapidly rotating low-mass stars
in close binaries that have radii consistent with evolutionary models
\citep{blake08}, and some binaries where one component has a consistent radius
and its companion is oversized \citep{kraus17}, implying that there are a
number of different factors that affect the over-inflation beyond enhanced
magnetic activity. Recent work from \citet{kesseli18} also shows that
neither rotation nor binarity is responsible for the inflated radii of
low-mass stars.

The number of precisely characterised low-mass stars is still low, due mainly
to their faintness. Pairs of eclipsing low-mass stars are still the benchmark
systems for such measurements \citep{lopezmorales07}, but few are known and
fewer still are bright enough to be studied at high precision. Moreover, the
effects of starspots on both stars makes modeling their light curves
complex. Low-mass stars in eclipsing binaries with more massive solar-type
stars are more numerous and brighter, but the large brightness contrast
between the two stars often means that the M star is essentially undetectable
spectroscopically, meaning that not only are these single-lined binaries
(making them less ideal for testing evolutionary models), but precise
temperature measurements for the M star are extremely
challenging. Interferometric studies of isolated low-mass stars can yield 
very precise radii, but lack the mass precision provided by binary
systems and are limited to a few nearby bright stars.

One type of system that is often overlooked is low-mass stars in detached
eclipsing binaries with white dwarfs. More than 3000 white dwarf plus
main-sequence star binaries are known \citep{rebassa16,ren18}, including more
than 70 eclipsing systems \citep{parsons15}. The small size of the white dwarf
(roughly Earth sized) results in very sharp eclipse features that can be used
to measure radii to very high precisions (1-2 per cent in the best cases, e.g.
\citealt{parsons10}). Moreover, in most cases both the white dwarf and
low-mass star are visible in optical spectra, making these double-lined 
binaries. Low-mass stars are roughly 10 times larger than white dwarfs,
meaning that the eclipse of the white dwarf is total and a clean spectrum of
the low-mass star can be obtained without contamination from the white
dwarf. Finally, the cooling of white dwarfs is well understood, making them
ideal objects for constraining the ages of their low-mass stellar companions.

It should be noted that these systems have experienced a brief common envelope
phase in their past evolution, when the progenitor star of the white dwarf
evolved off the main-sequence. During the common envelope phase (or rather
shortly prior to it) mass was transferred to the low-mass star. However, this
phase is extremely short ($10^3-10^4$\,years) compared to the thermal
timescale of a low-mass star ($10^8-10^9$\,years) and so has a negligible
effect on the star. The common envelope itself possesses much higher specific
entropy than the surface of the M dwarf, meaning that the star is thermally
isolated from the common envelope and hence essentially no accretion takes
place \citep{hjellming91}.

In this paper we present 16 high precision mass and radius measurements for M
dwarfs in eclipsing binaries with white dwarfs. Along with another 7
previously studied systems we also determine the effective temperatures,
metallicities and ages of these stars and compare these to the predictions of
evolutionary models.

\section{Observations and their reduction}

\begin{figure*}
  \begin{center}
    \includegraphics[width=0.95\textwidth]{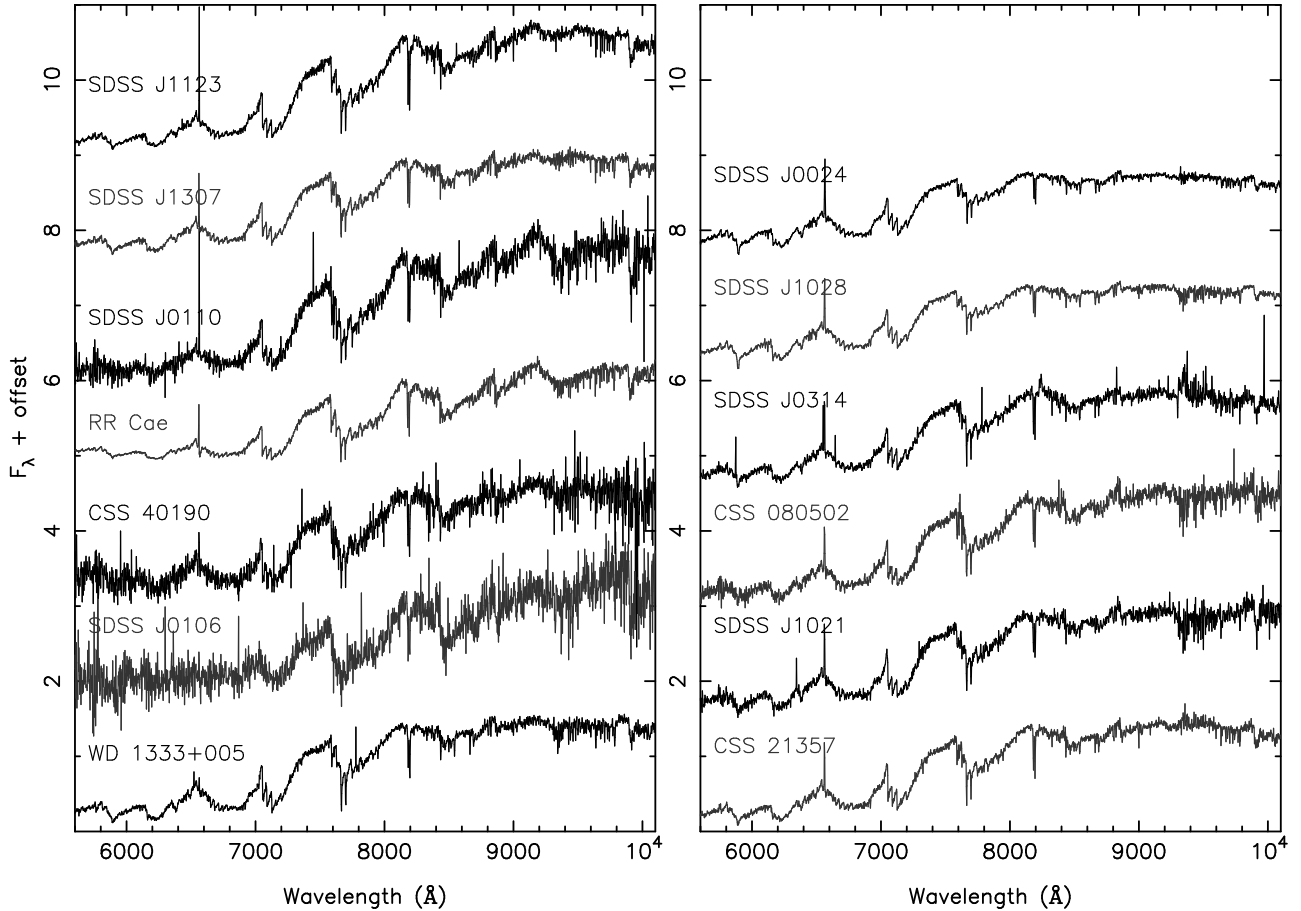}
    \caption{X-shooter VIS arm spectra of the M dwarfs in our binaries
      ordered from least massive (bottom-left) to most massive
      (top-right). These spectra were all obtained during the eclipse of the
      white dwarf and so only the M dwarf component is visible. The spectra
      have been binned by a factor of 10 for clarity. We do not plot the very
      noisy in-eclipse spectra of CSS\,09704, SDSS\,J1329+1230 or
      SDSS\,J2235+1428.} 
  \label{fig:specs}
  \end{center}
\end{figure*}

High cadence eclipse light curves for all our targets were obtained with the
high-speed frame-transfer cameras ULTRACAM \citep{dhillon07} on the 8.2m VLT
and 3.5m NTT in Chile and 4.2m WHT on La Palma and ULTRASPEC \citep{dhillon14}
on the 2.4m TNT in Thailand. Intermediate resolution optical and near-infrared
spectroscopy was obtained using the X-shooter spectrograph
\citep{vernet11} on the VLT, including observations of several M dwarf
spectral standard stars, which are detailed in Table~\ref{tab:standards}. All
of these observations and their reductions are detailed in \citet{parsons17}. In
addition to these data we also obtained a single X-shooter spectrum of the
system RR\,Cae on 30 August 2017. This was reduced in an identical manner to
the other X-shooter data. We plot the X-shooter VIS arm spectra taken during
the eclipse of the white dwarf (i.e. of the M dwarf component only) in
Figure~\ref{fig:specs}.

\section{Stellar parameters}

\subsection{Masses and radii}

\begin{figure}
  \begin{center}
    \includegraphics[width=\columnwidth]{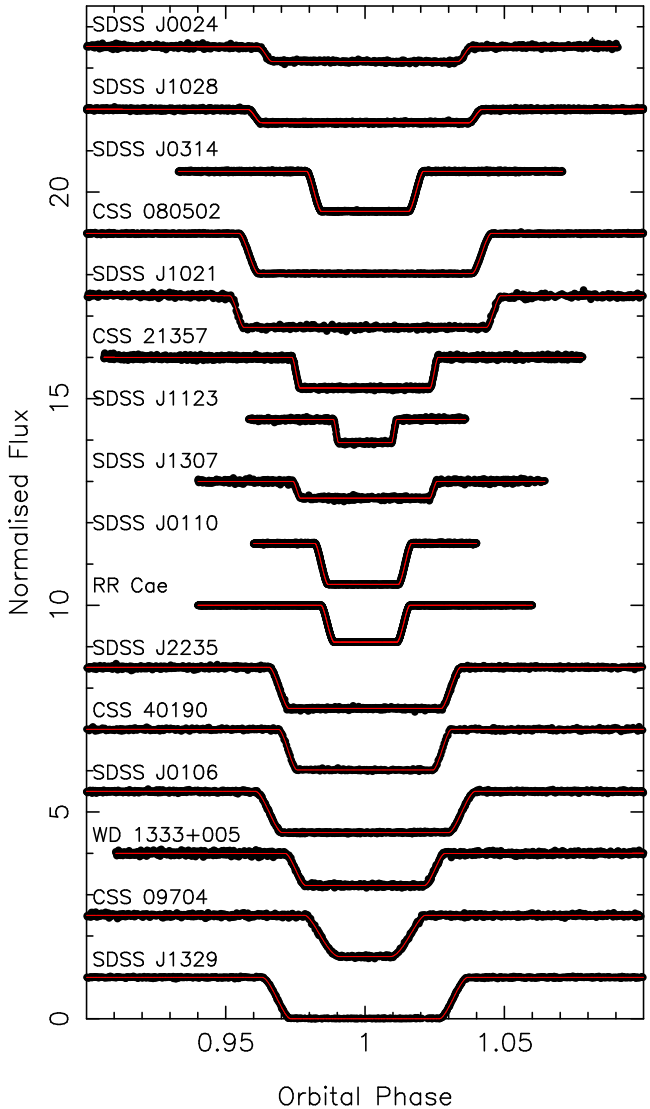}
    \caption{Light curves of the eclipse of the white dwarf in our
      binaries with model fits overplotted in red. Light curves are ordered
      from the least massive (bottom) to most massive (top) M star and are
      offset vertically by 1.5. The light curves shown are in the $g'$ band or
      the $KG5$ band (for WD\,1333+005, SDSS\,J1307+2156, SDSS\,J1123$-$1155,
      CSS\,21357 and SDSS\,J1028+0931). The $KG5$ filter is a broad band
      ($u'+g'+r'$) filter, see \citet{hardy17} for details.}
  \label{fig:eclipses}
  \end{center}
\end{figure}

\begin{figure*}
  \begin{center}
    \includegraphics[width=0.9\textwidth]{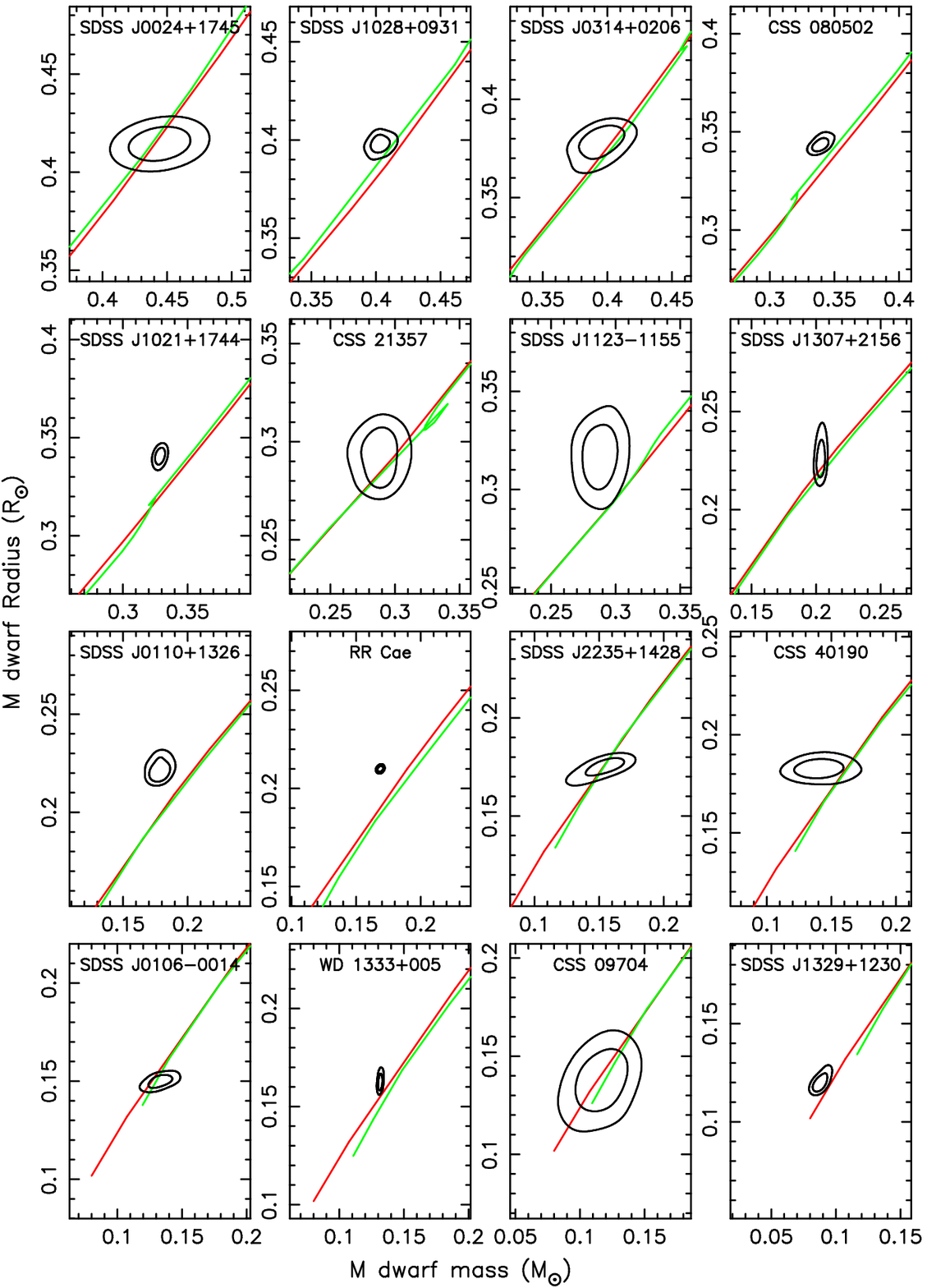}
    \caption{Constraints on the masses and radii (volume averaged) of the M
      dwarfs in all our systems, shown as contours (68 and 95 percentile
      regions). Red lines show the \citet{baraffe15} theoretical mass--radius
      relationship using the model with the closest match in age. Green lines
      show the \citet{dotter08} relationship with matching ages and
      metallicities (solar metallicity is used when we have no
      constraints). All plots are on the same scale and centred on the mean
      mass-radius value for each M dwarf and run from most massive (top-left)
      to least massive (bottom-right).}
  \label{fig:mr_conts}
  \end{center}
\end{figure*}

A detailed account of how we measured the masses and radii of each star in all
our binaries is given in \citet{parsons17}. In brief, radial velocity
semi-amplitudes were measured from our spectra and combined with fits to the
eclipse light curves (see Figure~\ref{fig:eclipses}). However, fits to the
white dwarf eclipse alone are degenerate (they only give the relationship
between the two radii as a function of inclination) and an additional piece of
information is required to determine the inclination and hence solve for the
physical parameters. This additional information could be: 1) the rotational
broadening of the M star, 2) the gravitational redshift of the white dwarf or,
3) the depth of the secondary eclipse (the transit of the white dwarf in front
of the M star). Generally, each system requires a different technique and we
refer readers to \citet{parsons17} for details of how each individual system
was solved. The masses and radii of the M dwarfs in our sample are listed
in Table~\ref{tab:mdparas} and shown in Figure~\ref{fig:mr_conts}. Note that
many of these M stars are tidally distorted by their white dwarf companions,
therefore, the quoted radii are the volume averaged radii of the stars, which
is most closely representative of the radii the stars would have if they
  were isolated.

\subsection{Spectral types}

\begin{table}
 \centering
  \caption{M dwarf template stars used in this study} 
  \label{tab:standards}
  \begin{tabular}{@{}lccc@{}}
  \hline
  Name                       & SpT   & $V$ & $K$ \\
  \hline
  LP\,887$-$70               & M1.0V & 11.00 &  7.12 \\
  LP\,905$-$56               & M1.5V & 11.22 &  6.99 \\
  GJ\,2066                   & M2.0V & 10.09 &  5.77 \\
  GJ\,588                    & M2.5V &  9.31 &  4.76 \\
  GJ\,812                    & M3.0V & 11.92 &  7.06 \\
  GJ\,849                    & M3.5V & 10.37 &  5.59 \\
  GJ\,876                    & M4.0V & 10.19 &  5.01 \\
  GJ\,3366                   & M4.5V & 14.54 &  9.18 \\
  GJ\,2045                   & M5.0V & 15.36 &  9.37 \\
  V645\,Cen                  & M5.5V & 11.13 &  4.38 \\
  Wolf\,359                  & M6.0V & 13.51 &  6.08 \\
  LP\,731$-$47               & M6.5V & 17.53 & 10.79 \\
  2MASS\,J01273195$-$3140031 & M7.0V & 20.40 & 11.66 \\
  LHS\,2021                  & M7.5V & 19.06 & 10.76 \\
  2MASS\,J10481463$-$3956062 & M8.0V & 17.53 &  8.45 \\
  \hline
\end{tabular}
\end{table}

The spectral types of the M dwarfs in our sample were determined using
template fitting. We observed a number of spectral type template stars with
X-shooter using an identical instrumental setup to our main science
observations. These template stars are detailed in Table~\ref{tab:standards}.
We used the spectra of our targets taken during the eclipse of the white dwarf
and fitted the \Ion{K}{i} 7699\,{\AA}, \Ion{Na}{i} 8183/8194\,{\AA} and
\Ion{K}{i} 1.252\,$\mu$m lines with each template spectrum. We artificially
broadened the lines of the template spectra to fit the observed lines, taking
into account any additional smearing of the lines from the velocity shift
during the exposure \citep[see][for details of this method]{marsh94}. We
applied a high-pass filter to both the observed and broadened template spectra
before comparing them in order to prevent the continuum dominating the
fit. The best fit template then yields both the rotational broadening of the M
star as well as its spectral type. 

In five cases the in-eclipse spectrum of the M dwarf was very poor and the
template fitting technique was not possible. In these cases we constrained the
spectral type of the M star using its $r-i$ colour measured from the eclipse
light curves and the relations of \citet{west05}. The spectral types of all
our M dwarfs are listed in Table~\ref{tab:mdparas}.

\subsection{Effective temperatures}

Effective temperatures were determined by comparing our in-eclipse M dwarf
spectra with a library of synthetic spectra. We used the BT-Settl model
spectra from \citet{allard13}, which are suitable for the low temperatures of
M dwarfs. The surface gravity was fixed to the nearest model based on the
measured mass and radius (either $\log{g}=4.5$ or $5.0$). The metallicity was
fixed at solar ([Fe/H]$=0$) which is the closest model value to our measured
metallicities (see Section~\ref{sec:meta}), and no
$\alpha$-element enhancement. We then used a grid of temperatures from 2300 to
4500\,K in steps of 100\,K. Each model was first degraded to match the
resolution of our X-shooter spectra and then rotationally broadened based on
the measured $v\sin{i}$ values. Models were compared over the X-shooter VIS
arm wavelengths (5600--10000\,{\AA}) excluding the region around the H$\alpha$
line (6500--6650\,{\AA}); note that our spectra were corrected for telluric
features (using observations of telluric standard stars taken shortly before
or after our science spectra). For each object the best-fitting model was
determined via $\chi^2$ minimization (see Table~\ref{tab:mdparas} for the
results).

In the case of objects with poor signal-to-noise ratio spectra the effective
temperature was instead determined by fitting the observed spectral energy
distribution (SED) of the binary. A model white dwarf spectrum was first
removed from the SED based on the fitted white dwarf parameters (see
\citealt{parsons17} for the white dwarf parameters), then the SED was fitted
using the virtual observatory SED analyzer (VOSA, \citealt{bayo08}). These
values are also listed in Table~\ref{tab:mdparas} and generally have a larger
uncertainty than the spectral fits. 

We also applied these fits to all of the previously published objects
listed in Table~\ref{tab:mdparas} using the data presented in those studies.

\begin{table*}
 \centering
  \caption{M dwarf mass-radius measurements obtained from detached, eclipsing
    binaries with white dwarfs. All temperature, metallicity and age
    measurements are from this paper. $\pi$ is the Gaia DR2 parallax. 
    Effective temperatures indicated with a $^{\dagger}$ symbol were determined
    from SED fitting rather than fitting with synthetic spectra. Spectral
    types indicated with a $^*$ symbol were determined from the in-eclipse 
    $r-i$ colour instead of template fitting. References for the other 
    measurements are given in the final column: (1) This paper, (2)
    \citet{parsons16}, (3) \citet{parsons12}, (4) \citet{pyrzas12}, (5)
    \citet{parsons12uc}, (6) \citet{parsons10}, (7) \citet{parsons12sp}.} 
  \label{tab:mdparas}
  \tabcolsep=0.13cm
  \begin{tabular}{@{}lcccclcclcc@{}}
  \hline
  Object             & $g$ mag & P$_\mathrm{orb}$ & Mass & Radius & $T_\mathrm{eff}$ & [Fe/H] & Age  & Sp type & $\pi$ & Ref \\
   & & (hrs) & (\MSUN) & (\RSUN) & (K) & (dex) & (Gyr) & & (mas) & \\
  \hline
  SDSS\,J0024$+$1745 & 18.71   & 4.801  & $0.444\pm0.016$ & $0.414\pm0.006$ & $3400\pm100$ & $-0.01\pm0.12$ & $2.81_{-0.26}^{+0.35}$ & M$2.5\pm0.5$ & $3.70\pm0.12$ & 1 \\
  SDSS\,J1028$+$0931 & 16.40   & 5.641  & $0.403\pm0.005$ & $0.398\pm0.003$ & $3500\pm100$ & $+0.04\pm0.12$ & $5.30_{-0.60}^{+0.85}$ & M$2.5\pm0.5$ & $5.70\pm0.06$ & 1 \\
  SDSS\,J0314$+$0206 & 16.95   & 7.327  & $0.395\pm0.012$ & $0.377\pm0.006$ & $3400\pm100$ & $-0.23\pm0.12$ & $0.90_{-0.06}^{+0.07}$ & M$3.0\pm1.0$ & $2.33\pm0.13$ & 1 \\
  QS\,Vir            & 14.66   & 3.618  & $0.382\pm0.006$ & $0.381\pm0.003$ & $3300\pm100$ & -              & $0.68_{-0.01}^{+0.02}$ & M$3.0\pm0.5$ & $19.96\pm0.06$ & 2 \\
  CSS\,080502        & 17.08   & 3.587  & $0.340\pm0.005$ & $0.344\pm0.003$ & $3300\pm100$ & $+0.02\pm0.12$ & $2.52_{-0.13}^{+0.15}$ & M$3.5\pm0.5$ & - & 1 \\
  SDSS\,J1021$+$1744 & 19.51   & 3.369  & $0.329\pm0.003$ & $0.340\pm0.003$ & $3300\pm100$ & $-0.14\pm0.12$ & $2.71_{-0.26}^{+0.31}$ & M$3.0\pm0.5$ & $2.28\pm0.44$ & 1 \\
  CSS\,21357         & 17.29   & 5.962  & $0.289\pm0.011$ & $0.293\pm0.009$ & $3300\pm100$ & $-0.19\pm0.12$ & $0.84_{-0.05}^{+0.06}$ & M$3.0\pm0.5$ & $6.08\pm0.13$ & 1 \\
  SDSS\,J1123$-$1155 & 17.99   & 18.459 & $0.288\pm0.009$ & $0.317\pm0.011$ & $3400\pm100$ & $+0.08\pm0.12$ & $2.02_{-0.18}^{+0.07}$ & M$3.5\pm0.5$ & $6.32\pm0.10$ & 1 \\
  SDSS\,J1212$-$0123 & 16.77   & 8.061  & $0.273\pm0.002$ & $0.306\pm0.007$ & $3300\pm100$ & $+0.00\pm0.12$ & $3.26_{-0.44}^{+0.46}$ & M$4.0\pm0.5$ & $4.80\pm0.11$ & 3 \\
  SDSS\,J1307$+$2156 & 18.25   & 5.192  & $0.204\pm0.002$ & $0.227\pm0.007$ & $3200\pm100$ & $-0.06\pm0.12$ & $1.88_{-0.06}^{+0.06}$ & M$4.0\pm0.5$ & $9.43\pm0.08$ & 1 \\
  SDSS\,J0110$+$1326 & 16.53   & 7.984  & $0.179\pm0.005$ & $0.222\pm0.004$ & $3200\pm100$ & $-0.03\pm0.12$ & $3.29_{-0.46}^{+0.58}$ & M$4.5\pm0.5$ & $3.60\pm0.10$ & 1 \\
  RR\,Cae            & 14.57   & 7.289  & $0.169\pm0.001$ & $0.210\pm0.001$ & $3100\pm100$ & $-0.35\pm0.12$ & $6.11_{-0.47}^{+0.65}$ & M$4.0\pm0.5$ & $47.16\pm0.02$ & 1 \\
  SDSS\,J1210$+$3347 & 16.94   & 2.988  & $0.158\pm0.006$ & $0.200\pm0.006$ & $3100\pm100$ & -              & $8.19_{-0.74}^{+0.93}$ & M$5.0\pm0.5$ & $21.69\pm0.08$ & 4 \\
  SDSS\,J2235$+$1428 & 18.59   & 3.467  & $0.151\pm0.013$ & $0.174\pm0.004$ & $3000\pm200^{\dagger}$ & -              & $5.45_{-1.63}^{+2.55}$ & M$5.0\pm1.0^*$ & - & 1 \\
  CSS\,40190         & 18.16   & 3.123  & $0.142\pm0.013$ & $0.183\pm0.003$ & $3100\pm100$ & -              & $2.62_{-0.22}^{+0.14}$ & M$5.0\pm0.5$ & $3.51\pm0.29$ & 1 \\
  SDSS\,J0106$-$0014 & 18.14   & 2.040  & $0.133\pm0.007$ & $0.150\pm0.002$ & $2900\pm150^{\dagger}$ & -              & $2.86_{-0.44}^{+0.65}$ & M$6.0\pm1.0^*$ & $3.16\pm0.26$ & 1 \\
  SDSS\,J0138$-$0016 & 18.84   & 1.746  & $0.132\pm0.004$ & $0.165\pm0.003$ & $3000\pm100$ & $-0.56\pm0.12$ & $11.13_{-0.39}^{+0.51}$ & M$5.0\pm0.5$ & $20.09\pm0.33$ & 5 \\
  WD\,1333+005       & 17.41   & 2.927  & $0.132\pm0.001$ & $0.163\pm0.003$ & $3100\pm100$ & $-0.25\pm0.12$ & $5.51_{-0.47}^{+0.58}$ & M$5.0\pm0.5$ & $10.63\pm0.12$ & 1 \\
  GK\,Vir            & 16.81   & 8.264  & $0.116\pm0.003$ & $0.146\pm0.003$ & $3000\pm200^{\dagger}$ & -              & $2.64_{-0.45}^{+0.59}$ & M$4.5\pm0.5$ & $2.11\pm0.13$ & 3 \\
  CSS\,09704         & 18.41   & 3.756  & $0.116\pm0.014$ & $0.137\pm0.011$ & $2900\pm250^{\dagger}$ & -              & $5.22_{-1.83}^{+3.50}$ & M$6.0\pm1.0^*$ & $1.93\pm0.40$ & 1 \\
  NN\,Ser            & 16.43   & 3.122  & $0.111\pm0.004$ & $0.141\pm0.002$ & $2900\pm200^{\dagger}$ & -              & $2.19_{-0.34}^{+0.50}$ & M$4.0\pm0.5$ & $1.92\pm0.10$ & 6 \\
  SDSS\,J1329$+$1230 & 17.26   & 1.943  & $0.088\pm0.004$ & $0.121\pm0.004$ & $2700\pm200^{\dagger}$ & -              & $7.42_{-2.15}^{+2.95}$ & M$8.0\pm1.0^*$ & $4.57\pm0.14$ & 1 \\
  SDSS\,J0857$+$0342 & 17.95   & 1.562  & $0.087\pm0.012$ & $0.104\pm0.004$ & $2600\pm300^{\dagger}$ & -              & $1.63_{-0.34}^{+0.56}$ & M$8.0\pm0.5^*$ & $0.99\pm0.30$ & 7 \\
  \hline
\end{tabular}
\end{table*}

\subsection{Metallicities} \label{sec:meta}

Metallicities of M dwarfs are notoriously difficult to measure and generally
rely on empirical relations derived from M dwarfs in binaries with higher mass
F, G and K stars. To measure the metallicities of our stars we used the
semi-empirical method outlined in \citet{newton14}, which relies solely on the
equivalent width of the \Ion{Na}{i} 2.205/2.209\,$\mu$m absorption doublet,
which is covered in our X-shooter data. This method has been shown to give
[Fe/H] values to (at best) a precision of up to 0.12 dex for M1--M5 dwarfs. It
has yielded reliable metallicities in other white dwarf plus M dwarf binaries
using similar X-shooter data \citep{rebassa16z}.

Our measured [Fe/H] values are detailed in Table~\ref{tab:mdparas} for those M
dwarfs with spectral types between M1 and M5. In several cases the quality of
the near-infrared spectrum was insufficient to give reliable equivalent width
measurements for the sodium doublet.

\subsection{Ages}

Ages are extremely difficult to measure for low-mass main-sequence
stars. While the ages of more massive solar-type stars can be constrained
using chromospheric activity indicators \citep[e.g.][]{soderblom91}, isochrone
fitting \citep{serenelli13} or gyrochronology \citep{barnes10}, 
the situation is more complicated for M dwarfs and most reliable measurements
come from low-mass stars in clusters. Given that the radius of a low-mass star
can change as much as 5 per cent between 1 and 10\,Gyr \citep{baraffe15}, it
is important to compare our measured radii with theoretical models of the
right age. Fortunately, the white dwarf companions to our M dwarfs can be used
to constrain the ages of our stars.

\begin{figure*}
  \begin{center}
    \includegraphics[width=0.95\textwidth]{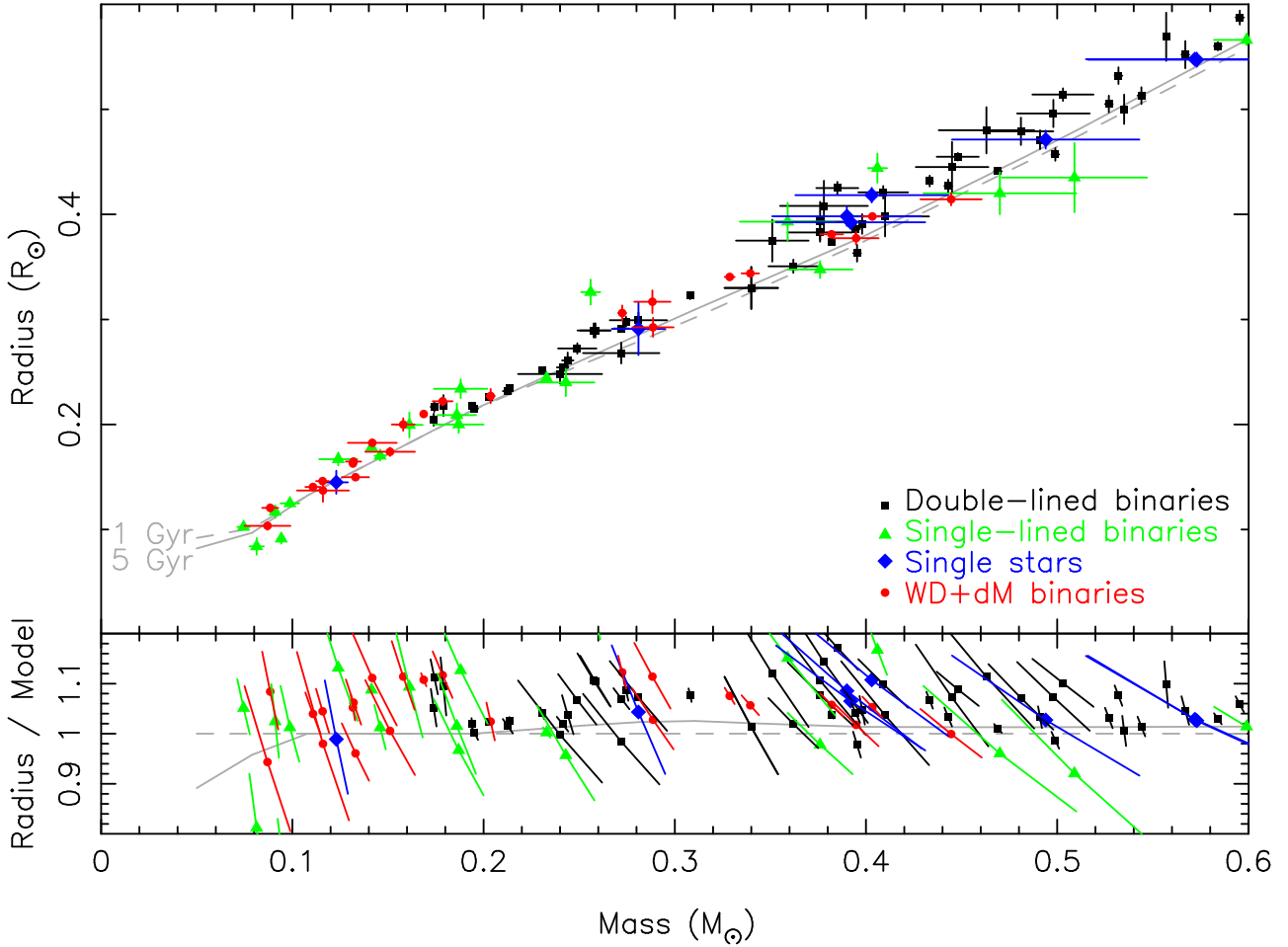}
    \caption{Mass-radius plot for low-mass stars (with mass and radius
      uncertainties of less than 10 per cent). The type of system that the
      measurement came from is indicated by the different colours and symbols
      and all are detailed in either Table~\ref{tab:mdparas} (red points) or
      the Appendix (all other points). Also shown are the theoretical
      mass-radius tracks from \citet{baraffe03,baraffe15}.}
  \label{fig:mdmr}
  \end{center}
\end{figure*}

Since the white dwarf parameters are known to high precision, their cooling
ages can be calculated to a few per cent. Normally an initial-to-final mass
relationship is then used to estimate the mass of the progenitor star of the
white dwarf \citep[e.g.][]{catalan08} and determine the main-sequence lifetime
and therefore establish the total age of the object. However, in the case of
close binary systems such as those presented here, this approach is not
appropriate because the evolution of the white dwarf progenitor was truncated
by its low-mass companion (due to a common envelope phase) and hence an
initial-to-final mass relation would under-predict the progenitor mass and
over-predict the total age. For these kinds of objects it is necessary to
properly reconstruct the evolutionary history of each system, which we did
following the algorithm described in detail by \citet[][section
3.2]{zorotovic11}. For a proper estimation of the errors, we have randomly
generated a Gaussian distribution for the white dwarf masses for each system,
centred on the observed mass and with a standard deviation that corresponds
to the measured error, as listed in \citet{parsons17}. We have then computed
the cooling ages and periods just after the common envelope phase for the
1000 masses for each system assuming disrupted magnetic braking
\citep{rappaport83}. These parameters were used to reconstruct their
evolutionary histories assuming a common envelope efficiency in the range of
0.2--0.3 and no contributions from recombination energy. The derived total age
of each system is listed in Table~\ref{tab:mdparas}, corresponding to the
median of all the possible solutions for each system, while the errors
represent the 34 percentile regions on each side of the median. We have also
verified that the errors in the effective temperatures of the white
dwarfs are negligible compared to the effect of the errors on the masses.

\section{Comparison to theoretical models}

\subsection{The mass-radius relation}

Table~\ref{tab:mdparas} lists all of our measurements as well as those of
other M dwarfs in eclipsing binaries with white dwarfs. The masses and radii
of these objects are shown in Figure~\ref{fig:mdmr} along with other
precise mass-radius measurements collected from other sources (these are
detailed in the Appendix). Both Figure~\ref{fig:mr_conts} and
Figure~\ref{fig:mdmr} show large amounts of scatter in the measured radii of
low-mass stars relative to theoretical predictions. In some cases the
measured radii are consistent with models, while in others the radii are more
than 10 per cent larger than expected. For example, the white dwarf plus M
dwarf binaries CSS\,21357 and SDSS\,J1123$-$1155 have M dwarfs with
essentially identical masses ($0.289\pm0.011$\,\MSUN\, and
$0.288\pm0.009$\,\MSUN) but their radii differ from each other by 9 per cent,
with CSS\,21357 having a radius consistent with models while
SDSS\,J1123$-$1155 is substantially oversized. This trend has been seen
before, for example the eclipsing binary PTFEB132.707$+$19.810 contains two
low-mass stars, one of which has a radius consistent with evolutionary models,
while the other is 20 per cent larger than expected \citep{kraus17}. 

\begin{figure*}
  \begin{center}
    \includegraphics[width=0.95\textwidth]{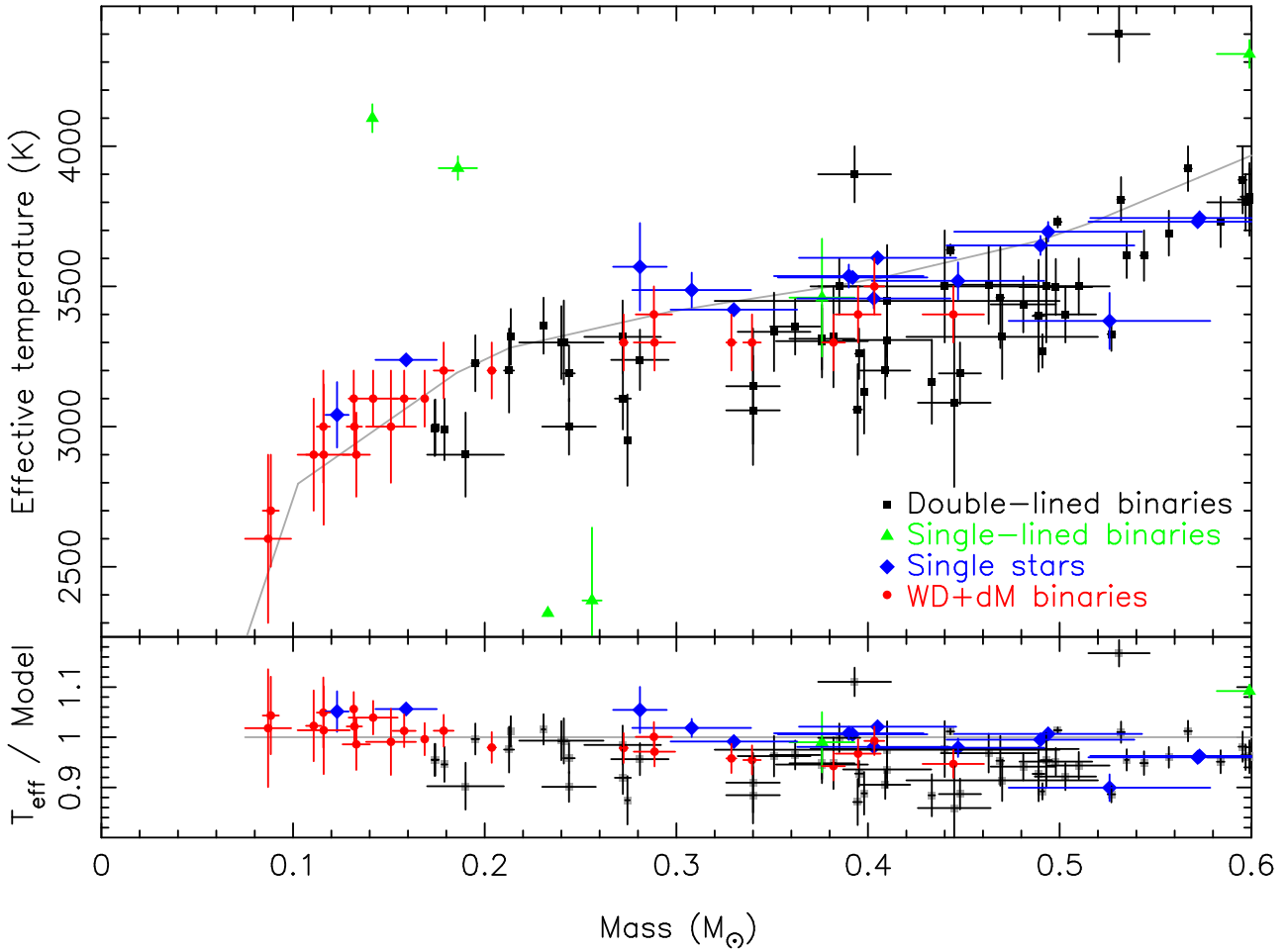}
    \caption{Mass-effective temperature plot for low-mass stars. The type of
      system that the measurement came from is indicated by the different
      colours and symbols and all are detailed in either
      Table~\ref{tab:mdparas} or the Appendix. Also shown are the 
      theoretical tracks from \citet{baraffe03,baraffe15} for an age of
      1\,Gyr.}
  \label{fig:mdmt}
  \end{center}
\end{figure*}

On average our measured radii are 6.2 per cent larger than predicted, although
the scatter on this value is substantial (4.8 per cent), indicating that this is
not a systematic offset. This over-inflation is seen in both partially
convective stars ($4.0\pm2.5$ per cent oversized) and fully convective stars
($7.1\pm5.1$ per cent oversized). Given that all of our stars are tidally
locked in short period ($P_\mathrm{orb}<1$\,d) binaries it is clear that this
over-inflation cannot be solely due to rapid rotation and enhanced magnetic
activity. Taking all of the measurements shown in Figure~\ref{fig:mdmr}
(i.e. all the values listed in Table~\ref{tab:mdparas} and the Appendix) gives
an average over-inflation of 5 per cent for both partially and fully convective
stars, but with a scatter of 5 per cent and little difference between the
different types of system. No clear difference is seen between the measured
radii of M dwarfs from different sources, confirming that the structure of M
dwarfs in close binaries with white dwarfs is not affected by common envelope
evolution.

\subsection{The mass-$T_\mathrm{eff}$ relation}

Figure~\ref{fig:mdmt} shows the measured temperatures of our M dwarfs as a
function of mass. Our systems mostly populate the low temperature end of the
plot, which is a region with few previous measurements. Interestingly, our
temperature measurements for the very low-mass stars ($<$0.2{\MSUN}) are in
agreement with theoretical predictions. The stars more massive than this are
all cooler than predicted by 100-200\,K, a trend seen in many stars in this
mass regime \citep{lopezmorales07}. On average, fully convective stars are
only 50\,K cooler than expected, while partially convective stars are 150\,K
cooler than theoretical models predict, however, there is not a clear boundary
between the two at 0.35{\MSUN}, rather the agreement with theoretical models
becomes better at lower masses. This behaviour is expected since the luminosity
of fully convective stars is set by the conditions in their very outermost
layers \citep{sirotkin10} and therefore their effective temperatures should
change little on expansion, as opposed to partially convective stars.

5 per cent of systems are significant outliers with temperatures more
than $\sim$500\,K hotter or cooler than models predict. It is unclear why
the temperatures of these specific objects are so discrepant, although we
note that the majority of these highly discrepant measurements are from
single-lined systems.

\subsection{The spectral type-$T_\mathrm{eff}$ relation}

\begin{figure}
  \begin{center}
    \includegraphics[width=\columnwidth]{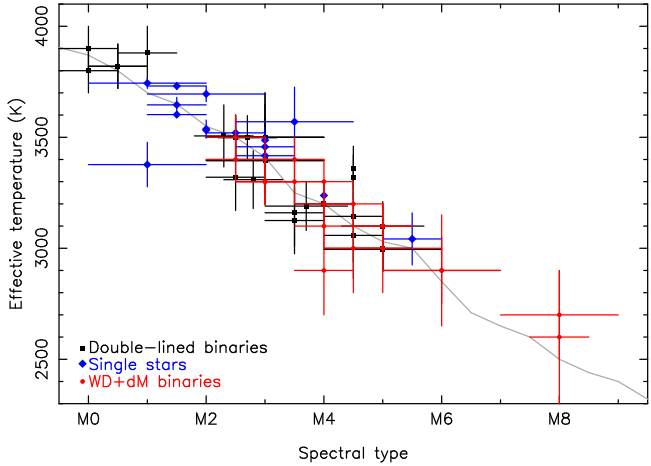}
    \caption{The spectral type-$T_\mathrm{eff}$ relation for M dwarfs. The
      type of system that the measurement came from is indicated by the
      different colours and symbols and all are detailed in
      Table~\ref{tab:mdparas} or the Appendix. The grey line is the observed
      relation from \citet{pecaut13}.}
  \label{fig:sptp}
  \end{center}
\end{figure}

Figure~\ref{fig:sptp} shows the spectral type-$T_\mathrm{eff}$ relation using
the stars in the sample with spectral type and temperature measurements listed
in Table~\ref{tab:mdparas} and the Appendix. Our new measurements show
excellent agreement with the empirical relation of \citet{pecaut13} (which
uses single stars), with all our measurements consistent to within 2 sigma.

\subsection{The mass-M$_{K_s}$ relation}

\begin{figure}
  \begin{center}
    \includegraphics[width=\columnwidth]{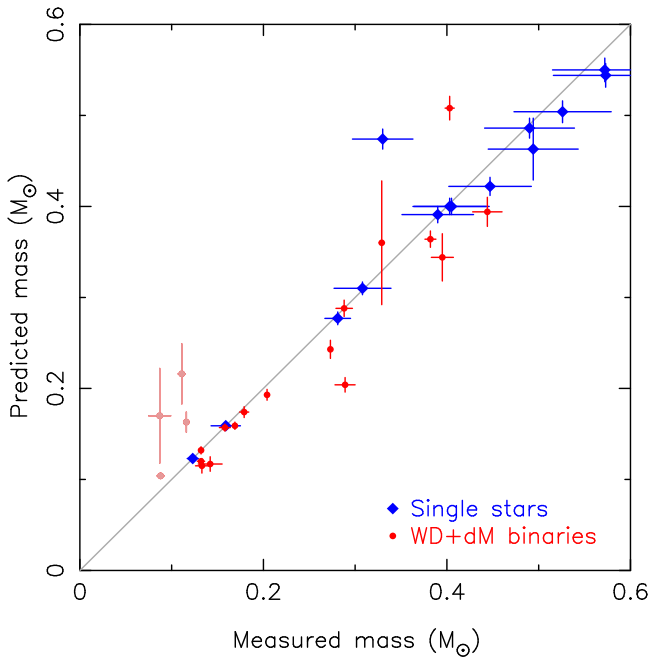}
    \caption{The measured masses of M dwarfs compared to those derived using
      the empirical mass-M$_{K_s}$ relation from \citet{mann18}. Faded red
      points are binaries where the white dwarf still contributes a
      substantial amount of the flux in the $K_s$ band and therefore may not
      have reliable predicted masses.}
  \label{fig:lum}
  \end{center}
\end{figure}

We determined the absolute $K_s$ band magnitudes of the M dwarfs in our
binaries using their 2MASS measurements (or UKIDSS for the fainter objects)
and parallaxes from Gaia data release 2 \citep{gaia18}, and list them in 
Table~\ref{tab:mdparas}. Two of our binaries
lack reliable Gaia parallaxes (CSS\,080502 and SDSS\,J2235+1428), while one
target has no near-infrared magnitude measurements (CSS\,09704), so we exclude
these three targets from our subsequent analysis. The remaining targets all
have high precision Gaia parallaxes (parallax/error$>$10), so their
distances can be determined by simple parallax inversion without significant
loss of accuracy \citep{bailer18} and hence absolute $K_s$ band magnitudes
can be derived. We used these values and the empirical mass-M$_{K_s}$ 
relationship from \citet{mann18} to estimate the masses of our M
dwarfs\footnote{\url{https://github.com/awmann/M_-M_K-}} and compared these to
our measured values. Figure~\ref{fig:lum} shows the difference between the
measured and predicted values for the M dwarfs in our binaries as well as the
single stars listed in the Appendix. We do not include M dwarfs in either
double- or single-lined binaries since measuring the $K_s$ magnitude for the M
dwarf alone is difficult in these binaries due to contamination from their
companion stars. This is one of the advantages of using white dwarf-M dwarf
binaries since in the majority of cases the M dwarf completely dominates the
flux in the $K_s$ band. However, this is
not always the case; in systems with very low mass M dwarfs and/or extremely
hot white dwarfs a significant amount of the $K_s$ band flux originates from
the white dwarf. We have highlighted these systems in Figure~\ref{fig:lum} and
the predicted masses of these objects has clearly been overestimated - these
points should be disregarded when comparing values in Figure~\ref{fig:lum}.
In-eclipse $K_s$ band measurements would remove this issue since the white
dwarf contribution would be obscured.

The mass-luminosity relationship systematically under predicts our measured
masses by 5--10 per cent. Single stars are more consistent, although their
masses are also slightly under predicted (albeit with larger mass errors).
This may be due to the enhanced number of star spots on M dwarfs in our close
binaries, which would lead to slightly fainter stars and therefore an
under prediction of their masses, although this should not be a large effect in
the $K_s$ band.

\subsection{The effect of age}

\begin{figure}
  \begin{center}
    \includegraphics[width=\columnwidth]{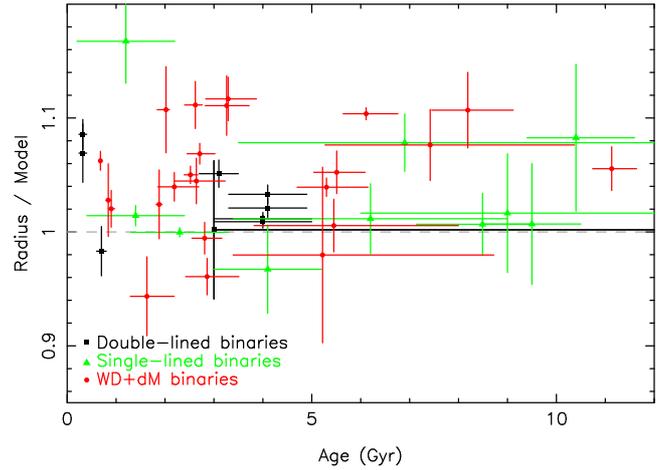}
    \caption{The over-inflation of M dwarfs as a function of their
      ages. Measured radii were compared to the closest \citet{baraffe15}
      model in age.}
  \label{fig:age}
  \end{center}
\end{figure}

The radii of low-mass stars change slowly over time, most notably in the first
billion years as they evolve on to the main-sequence. Once on the
main-sequence however, the radii of M dwarfs barely change over a Hubble
time. While these evolutionary effects are taken into account in theoretical
models, the lack of reliable age measurements for M dwarfs has prevented any
detailed investigation of the consistency of these models over a wide range of
ages. A difference in the measured over-inflation of young and old M dwarfs
could help reveal why evolutionary models consistently under-predict the radii
of low-mass stars.

Figure~\ref{fig:age} shows the over-inflation of M dwarfs as a function of
their total age (where the over-inflation is measured using the nearest model
in age). The white dwarf plus M dwarf binaries clearly cover a wide age range. 
However, there is no clear indication of more or less over-inflation as a
function of age. More objects with reliable age measurements, particularly old
objects ($>$7\,Gyr) will be needed to test this more robustly.

\subsection{The effect of metallicity}

Metallicity is expected to have a small but noticeable impact on the radius of
a low-mass star. Lower metallicities lead to decreased opacity of the outer
layers of the star and hence decreased radiation pressure, resulting in a
smaller star \citep[e.g.][]{chabrier97}. For example, for the theoretical
models of \citet{dotter08}, the difference in radius between a 0.2{\MSUN} star
with [Fe/H]=--1.0 and [Fe/H]=+0.5 is 9 per cent, which is detectable given our
precision. 

\begin{figure}
  \begin{center}
    \includegraphics[width=\columnwidth]{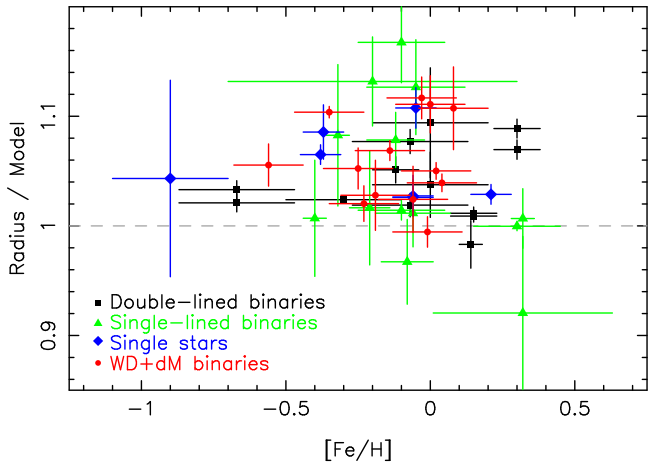}
    \caption{The over-inflation of M dwarfs as a function of their
      metallicity. Radii were compared to solar metallicity models from
      \citet{baraffe15}.}
  \label{fig:metal}
  \end{center}
\end{figure}

Figure~\ref{fig:metal} shows the over-inflation of M dwarfs as a function of
their metallicity. There is no clear evidence of metal-poor stars being
smaller or metal-rich stars being larger in this plot. However, the vast
majority of objects plotted in Figure~\ref{fig:metal} have roughly solar
metallicity and the extremes of metallicity are poorly sampled. For the white
dwarf plus M dwarf systems this is primarily because we determined the
metallicities using the semi-empirical method of \citet{newton14}, which is
only valid between -0.6$<$[Fe/H]$<$0.3. Its also worth noting that there has
been some criticism of metallicity calibrators based on the spectral analysis
of M dwarfs \citep[e.g.][]{lindgren17}, so Figure~\ref{fig:metal} should be
interpreted with some caution.

\subsection{The effect of rotation}

The effects of rotation on the radii of low mass stars was investigated in
detail by \citet{kraus11}, who found that M dwarfs in short period binaries
(P$_\mathrm{orb} < 1$\,d) were more oversized than those in longer period
systems. This is consistent with the theory that rapid rotation leads to
enhanced magnetic activity which inhibits convection leading to
inflation. 

\begin{figure}
  \begin{center}
    \includegraphics[width=\columnwidth]{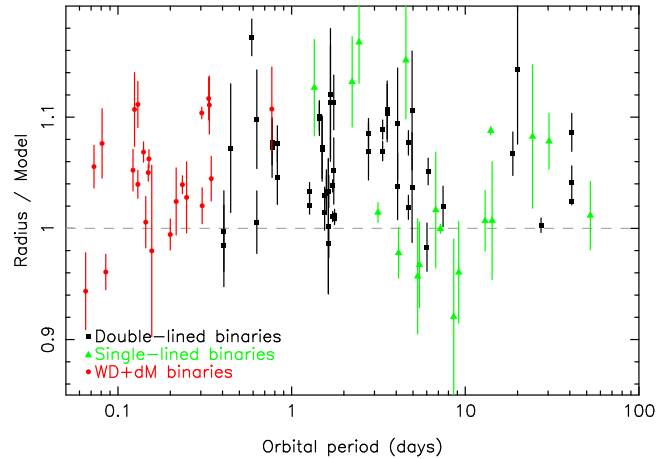}
    \caption{The over-inflation of M dwarfs as a function of their orbital
      period. Note that many stars in systems with periods longer than a few
      days are not synchronously rotating (usually rotating slower than the
      binary period) and generally have moderately eccentric orbits.}
  \label{fig:period}
  \end{center}
\end{figure}

In Figure~\ref{fig:period} we plot the over-inflation as a function of orbital
period. Our new systems all occupy the very shortest period end, extending
precision measurements to the shortest periods measured to date. While we
expect the M stars in our binaries to be tidally locked to the white dwarf and
hence have rotation periods equal to the orbital periods, that is not the case
for many other types of binary, particularly those with periods longer than a
few days. In these cases the stars generally rotate slower than the orbital
period. Furthermore, our new binaries all have very circular orbits (for
example the eccentricity of NN\,Ser has been constrained to $e<10^{-3}$,
\citealt{parsons14nnser}), but many of the longer period main-sequence
binaries have substantial eccentricities. These are listed in the
Appendix. 

In contrast to the results of \citet{kraus11}, Figure~\ref{fig:period} shows
little difference in the over-inflation of low-mass stars rotating faster or
slower than 1 day, although in both cases there is substantial scatter in the
radii. However, at periods longer than roughly 5 days the measured radii do
appear to be more consistent with theoretical predictions although they
  show a similar amount of scatter.

\section{Conclusions}

We have presented high-precision mass, radius, effective temperature and age
measurements for 23 M dwarfs in eclipsing binaries with white dwarfs, 16 of
which are new results. We have also determined the metallicities for 13 of these
objects. On average the radii of these stars are $6.2\pm4.8$ per cent larger
than theoretical models predict, although they show a large amount of scatter,
and around a quarter of them have measured radii consistent with models. No
difference is seen between partially and fully convective stars. The fact that
all of these stars are rapid rotators means that enhanced activity leading to
the suppression of convection cannot be the only cause of the discrepancy in
the radii of low-mass stars. 

We find that the measured temperatures of very low-mass M dwarfs ($< 0.2$\MSUN)
are in agreement with theoretical models, but more massive stars are
systematically cooler than models predict by $\sim$100\,K.

Finally, we find no clear trend in the over-inflation of M dwarfs as a
function of age or metallicity, but do find that M dwarfs rotating slower than
$\sim$5 days have on average radii more consistent with models, although there
is a similar amount of scatter compared to more rapidly rotating M dwarfs.

The results presented in this paper demonstrate the difficulty in determining
reliable parameters for low mass stars and by extension any planets that they
may host. The use of theoretical or empirical relations may still lead to
errors of 5--10 per cent in the radii of exoplanets around M dwarfs, generally
insufficient to constrain their internal structure and bulk composition.

\section*{Acknowledgements}

SGP acknowledges the support of the Leverhulme Trust. The research leading to
these results has received funding from the European Research Council under
the European Union's Seventh Framework Programme (FP/2007-2013) / ERC Grant
Agreement numbers 340040 (HiPERCAM) and 320964 (WDTracer). ULTRACAM, TRM, VSD,
and SPL are supported by the Science and Technology Facilities Council
(STFC). ARM acknowledges financial support from the MINECO Ram\'on y Cajal
programme RYC-2016-20254 and the grant AYA\-2017-86274-P, and from the AGAUR
(SGR-661/2017). Support for this work was provided by NASA through Hubble 
Fellowship grant \#HST-HF2-51357.001-A. MRS thanks for support from FONDECYT
(1141269) and Millennium Science Initiative, Chilean ministry of Economy:
Nucleus P10-022-F. MZ acknowledges support from CONICYT PAI (Concurso Nacional
de Inserci\'on en la Academia 2017, Folio 79170121) and CONICYT/FONDECYT
(Programa de Iniciaci\'on, Folio 11170559). This work has made use of data
obtained at the Thai National Observatory on Doi Inthanon, operated by
NARIT. The results presented in this paper are based on observations collected
at the European Southern Observatory under programme IDs 086.D-0161,
086.D-0265, 192.D-0270 and 099.D-0252.

\bibliographystyle{mnras}
\bibliography{md_mr}

\appendix

\section{A catalogue of well characterised M dwarfs}

\begin{table*}
 \centering
  \caption{M dwarfs with well constrained physical parameters. $e$ is the
    eccentricity of the orbit. References: (1) \citet{birkby12}, (2)
    \citet{blake08}, (3) \citet{carter11}, (4) \citet{creevey05}, (5)
    \citet{hartman11}, (6) \citet{hebb06}, (7) \citet{helminiak12}, (8)
    \citet{irwin09}, (9) \citet{irwin11}, (10) \citet{kraus11}, (11)
    \citet{lopezmorales07}, (12) \citet{vaccaro07}, (13) \citet{zhou15}, (14)
    \citet{dittmann17}, (15) \citet{hartman18}, (16) \citet{kraus17}, (17)
    \citet{cruz18}, (18) \citet{casewell18}, (19) \citet{doyle11}, (20)
    \citet{orosz12a}, (21) \citet{orosz12b}, (22) \citet{schwamb13}, (23)
    \citet{welsh15}, (24) \citet{pont06}, (25) \citet{beatty07}, (26) 
    \citet{diaz14}, (27) \citet{shporer17}, (28) \citet{chaturvedi18}, (29)
    \citet{ofir12}, (30) \citet{gomezchew14}, (31) \citet{eigmuller16}, (32)
    \citet{iglesias17}, (33) \citet{boetticher17}, (34) \citet{triaud13},
    (35) \citet{eigmuller18}, (36) \citet{mann13}.}
  \label{tab:mdwarfs}
  \tabcolsep=0.16cm
  \begin{tabular}{@{}lccccccccc@{}}
  \hline
  Name               & Mass              & Radius            &  $T_\mathrm{eff}$ & [Fe/H]         & Age           & Sp type     &  P$_\mathrm{orb}$ & $e$    & Ref \\
                     & (\MSUN)           & (\RSUN)           & (K)             & (dex)          & (Gyr)         &              & (days)          &        &     \\
  \hline
  \multicolumn{2}{l}{\bf Double lined eclipsing binaries:}\\
  19b-2-01387a       & $0.498\pm0.019$   & $0.496\pm0.013$   & $3498\pm100$    & -              & -             & M$2.7\pm0.5$ & 1.4985          & 0.01  & 1   \\
  19b-2-01387b       & $0.481\pm0.017$   & $0.479\pm0.013$   & $3436\pm100$    & -              & -             & -            & 1.4985          & 0.01  & 1   \\
  19c-3-01405a       & $0.410\pm0.023$   & $0.398\pm0.019$   & $3309\pm130$    & -              & -             & M$2.8\pm0.5$ & 4.9391          & 0.01  & 1   \\
  19c-3-01405b       & $0.376\pm0.024$   & $0.393\pm0.019$   & $3305\pm130$    & -              & -             & -            & 4.9391          & 0.01  & 1   \\
  19e-3-08413a       & $0.463\pm0.025$   & $0.480\pm0.022$   & $3506\pm140$    & -              & -             & M$2.3\pm0.5$ & 1.6734          & 0.01  & 1   \\
  19e-3-08413b       & $0.351\pm0.019$   & $0.375\pm0.020$   & $3338\pm140$    & -              & -             & -            & 1.6734          & 0.01  & 1   \\
  SDSS\,J0318-0100a  & $0.272\pm0.020$   & $0.268\pm0.010$   & $3320\pm130$    & -              & -             & -            & 0.4070          & 0.00  & 2   \\
  SDSS\,J0318-0100b  & $0.240\pm0.022$   & $0.248\pm0.009$   & $3300\pm130$    & -              & -             & -            & 0.4070          & 0.00  & 2   \\
  KOI-126a           & $0.2413\pm0.0030$ & $0.2543\pm0.0014$ & $3300\pm150$    & $+0.15\pm0.08$ & $4.0\pm1.0$   & -            & 1.7671          & 0.02  & 3   \\
  KOI-126b           & $0.2127\pm0.0026$ & $0.2318\pm0.0013$ & $3200\pm150$    & $+0.15\pm0.08$ & $4.0\pm1.0$   & -            & 1.7671          & 0.02  & 3   \\
  TrES-Her0-07621a   & $0.493\pm0.003$   & $0.453\pm0.060$   & $3500\pm150$    & -              & -             & M$3.0\pm1.0$ & 1.1208          & 0.00  & 4   \\
  TrES-Her0-07621b   & $0.489\pm0.003$   & $0.452\pm0.050$   & $3395\pm150$    & -              & -             & M$3.0\pm1.0$ & 1.1208          & 0.00  & 4   \\
  1RXS\,J1547+4508a  & $0.2576\pm0.0085$ & $0.2895\pm0.0068$ & -               & -              & -             & -            & 3.5500          & 0.00  & 5   \\
  1RXS\,J1547+4508b  & $0.2585\pm0.0080$ & $0.2895\pm0.0068$ & -               & -              & -             & -            & 3.5500          & 0.00  & 5   \\
  2MASS\,J0446+1901a & $0.470\pm0.050$   & $0.570\pm0.020$   & $3320\pm150$    & -              & $0.15\pm0.05$ & M$2.5\pm0.5$ & 0.6188          & 0.00  & 6   \\
  2MASS\,J0446+1901b & $0.190\pm0.020$   & $0.210\pm0.010$   & $2900\pm150$    & -              & $0.15\pm0.05$ & -            & 0.6188          & 0.00  & 6   \\
  ASAS\,J0113-3821a  & $0.612\pm0.030$   & $0.596\pm0.020$   & $3750\pm250$    & -              & -             & -            & 0.4456          & 0.00  & 7   \\
  ASAS\,J0113-3821b  & $0.445\pm0.019$   & $0.445\pm0.024$   & $3085\pm300$    & -              & -             & -            & 0.4456          & 0.00  & 7   \\
  GJ\,3236a          & $0.376\pm0.017$   & $0.3828\pm0.0072$ & $3313\pm110$    & -              & -             & -            & 0.7713          & 0.00  & 8   \\
  GJ\,3236b          & $0.281\pm0.015$   & $0.2992\pm0.0075$ & $3238\pm108$    & -              & -             & -            & 0.7713          & 0.00  & 8   \\
  LSPM\,J1112+7626a  & $0.3946\pm0.0023$ & $0.3860\pm0.0054$ & $3061\pm162$    & -              & -             & -            & 41.032          & 0.24  & 9   \\
  LSPM\,J1112+7626b  & $0.2745\pm0.0012$ & $0.2978\pm0.0047$ & $2952\pm163$    & -              & -             & -            & 41.032          & 0.24  & 9   \\
  MG1-78457a         & $0.527\pm0.002$   & $0.505\pm0.008$   & $3330\pm60$     & -              & -             & -            & 1.5862          & 0.00  & 10  \\
  MG1-78457b         & $0.491\pm0.001$   & $0.471\pm0.009$   & $3270\pm60$     & -              & -             & -            & 1.5862          & 0.00  & 10  \\
  MG1-116309a        & $0.567\pm0.002$   & $0.552\pm0.013$   & $3920\pm80$     & -              & -             & -            & 0.8271          & 0.00  & 10  \\
  MG1-116309b        & $0.532\pm0.002$   & $0.532\pm0.008$   & $3810\pm80$     & -              & -             & -            & 0.8271          & 0.00  & 10  \\
  MG1-506664a        & $0.584\pm0.002$   & $0.560\pm0.004$   & $3730\pm90$     & -              & -             & -            & 1.5485          & 0.00  & 10  \\
  MG1-506664b        & $0.544\pm0.002$   & $0.513\pm0.008$   & $3610\pm90$     & -              & -             & -            & 1.5485          & 0.00  & 10  \\
  MG1-646680a        & $0.499\pm0.002$   & $0.457\pm0.006$   & $3730\pm20$     & -              & -             & -            & 1.6375          & 0.00  & 10  \\
  MG1-646680b        & $0.443\pm0.002$   & $0.427\pm0.006$   & $3630\pm20$     & -              & -             & -            & 1.6375          & 0.00  & 10  \\
  MG1-1819499a       & $0.557\pm0.001$   & $0.569\pm0.023$   & $3690\pm80$     & -              & -             & -            & 0.6303          & 0.00  & 10  \\
  MG1-1819499b       & $0.535\pm0.001$   & $0.500\pm0.014$   & $3610\pm80$     & -              & -             & -            & 0.6303          & 0.00  & 10  \\
  MG1-2056316a       & $0.469\pm0.002$   & $0.441\pm0.002$   & $3460\pm180$    & -              & -             & -            & 1.7228          & 0.00  & 10  \\
  MG1-2056316b       & $0.382\pm0.001$   & $0.374\pm0.002$   & $3320\pm180$    & -              & -             & -            & 1.7228          & 0.00  & 10  \\
  CM\,Dra\,a         & $0.2307\pm0.0010$ & $0.2516\pm0.0020$ & $3360\pm100$    & $-0.67\pm0.20$ & $4.10\pm0.80$ & M$4.5$       & 1.2684          & 0.01  & 11  \\
  CM\,Dra\,b         & $0.2136\pm0.0001$ & $0.2347\pm0.0019$ & $3320\pm100$    & $-0.67\pm0.20$ & $4.10\pm0.80$ & M$4.5$       & 1.2684          & 0.01  & 11  \\
  YY\,Gem\,a         & $0.5992\pm0.0047$ & $0.6191\pm0.0057$ & $3820\pm100$    & $+0.10\pm0.20$ & $0.32\pm0.08$ & M$0.5\pm0.5$ & 0.8143          & 0.00  & 11  \\
  YY\,Gem\,b         & $0.5992\pm0.0047$ & $0.6191\pm0.0057$ & $3820\pm100$    & $+0.10\pm0.20$ & $0.32\pm0.08$ & M$0.5\pm0.5$ & 0.8143          & 0.00  & 11  \\
  CU\,Cnc\,a         & $0.4333\pm0.0017$ & $0.4317\pm0.0052$ & $3160\pm150$    & -              & $0.32\pm0.08$ & M$3.5\pm0.5$ & 2.7715          & 0.00  & 11  \\
  CU\,Cnc\,b         & $0.3980\pm0.0014$ & $0.3908\pm0.0094$ & $3125\pm150$    & -              & $0.32\pm0.08$ & M$3.5\pm0.5$ & 2.7715          & 0.00  & 11  \\
  GU\,Boo\,a         & $0.610\pm0.007$   & $0.623\pm0.016$   & $3920\pm130$    & -              & -             & -            & 0.4887          & 0.00  & 11  \\
  GU\,Boo\,b         & $0.599\pm0.006$   & $0.620\pm0.020$   & $3810\pm130$    & -              & -             & -            & 0.4887          & 0.00  & 11  \\
  OGLE\,BW3\,V38a    & $0.44\pm0.07$     & $0.51\pm0.04$     & $3500\pm200$    & -              & -             & M$3.0\pm1.0$ & 0.1984          & 0.00  & 11  \\
  OGLE\,BW3\,V38b    & $0.41\pm0.09$     & $0.44\pm0.06$     & $3448\pm200$    & -              & -             & -            & 0.1984          & 0.00  & 11  \\
  TRES-Her0-07621a   & $0.493\pm0.003$   & $0.453\pm0.060$   & $3500\pm200$    & -              & -             & M$3.0\pm1.0$ & 1.1208          & 0.00  & 11  \\
  TRES-Her0-07621b   & $0.489\pm0.003$   & $0.452\pm0.050$   & $3395\pm200$    & -              & -             & M$3.0\pm1.0$ & 1.1208          & 0.00  & 11  \\
  UNSW-TR\,2a        & $0.529\pm0.035$   & $0.641\pm0.050$   & -               & -              & -             & -            & 2.1167          & 0.00  & 11  \\
  UNSW-TR\,2b        & $0.512\pm0.035$   & $0.608\pm0.060$   & -               & -              & -             & -            & 2.1167          & 0.00  & 11  \\
  LP\,133-373a       & $0.340\pm0.014$   & $0.33\pm0.02$     & $3058\pm195$    & -              & $>$3.0        & M$4.5\pm0.5$ & 1.6280          & 0.00  & 12  \\
  LP\,133-373b       & $0.340\pm0.014$   & $0.33\pm0.02$     & $3144\pm206$    & -              & $>$3.0        & M$4.5\pm0.5$ & 1.6280          & 0.00  & 12  \\
  HATS551-027a       & $0.2440\pm0.0030$ & $0.2610\pm0.0075$ & $3190\pm100$    & $+0.00\pm0.20$ & -             & -            & 4.0770          & 0.00  & 13  \\
  HATS551-027b       & $0.1790\pm0.0015$ & $0.2180\pm0.0100$ & $2990\pm110$    & $+0.00\pm0.20$ & -             & -            & 4.0770          & 0.00  & 13  \\
  LP\,661-13a        & $0.3080\pm0.0008$ & $0.3226\pm0.0033$ & -               & $-0.07\pm0.20$ & -             & M$3.5\pm0.5$ & 4.7044          & 0.00  & 14  \\
  LP\,661-13b        & $0.1940\pm0.0003$ & $0.2174\pm0.0023$ & -               & $-0.07\pm0.20$ & -             & M$4.3\pm0.5$ & 4.7044          & 0.00  & 14  \\
  \hline
  \multicolumn{10}{r}{\it continues on the next page...}\\
\end{tabular}
\end{table*}

\begin{table*}
 \centering
  \tabcolsep=0.16cm
  \begin{tabular}{@{}lccccccccc@{}}
  \hline
  Name               & Mass              & Radius            &  $T_\mathrm{eff}$ & [Fe/H]         & Age           & Sp type     &  P$_\mathrm{orb}$ & $e$    & Ref \\
                     & (\MSUN)           & (\RSUN)           & (K)             & (dex)          & (Gyr)         &              & (days)          &        &     \\
  \hline
  HAT-TR-318-007a    & $0.4480\pm0.0110$ & $0.4548\pm0.0036$ & $3190\pm110$    & $+0.30\pm0.08$ & -             & M$3.7\pm0.7$ & 3.3440          & 0.01  & 15  \\
  HAT-TR-318-007b    & $0.2721\pm0.0042$ & $0.2913\pm0.0024$ & $3100\pm110$    & $+0.30\pm0.08$ & -             & M$5.0\pm0.7$ & 3.3440          & 0.01  & 15  \\
  PTFEB\,132+19a     & $0.3953\pm0.0020$ & $0.363\pm0.008$   & $3260\pm90$     & $+0.14\pm0.04$ & $0.60-0.80$   & -            & 6.0157          & 0.00  & 16  \\
  PTFEB\,132+19b     & $0.2098\pm0.0014$ & $0.272\pm0.012$   & $3120\pm110$    & $+0.14\pm0.04$ & $0.60-0.80$   & -            & 6.0157          & 0.00  & 16  \\
  17e-3-02003a       & $0.597\pm0.020$   & $0.611\pm0.095$   & $3800\pm100$    & -              & -             & $0.0\pm0.5$  & 1.2250          & 0.00  & 17  \\
  17e-3-02003b       & $0.510\pm0.016$   & $0.540\pm0.110$   & $3500\pm100$    & -              & -             & $2.5\pm0.5$  & 1.2250          & 0.00  & 17  \\
  17h-4-01429a       & $0.503\pm0.016$   & $0.514\pm0.006$   & $3400\pm100$    & -              & -             & $3.0\pm0.5$  & 1.4446          & 0.00  & 17  \\
  17h-4-01429b       & $0.409\pm0.013$   & $0.421\pm0.006$   & $3200\pm100$    & -              & -             & $4.0\pm0.5$  & 1.4446          & 0.00  & 17  \\
  19c-3-08647a       & $0.393\pm0.019$   & $0.494\pm0.069$   & $3900\pm100$    & -              & -             & M$0.0\pm0.5$ & 0.8675          & 0.00  & 17  \\
  19c-3-08647b       & $0.244\pm0.014$   & $0.422\pm0.077$   & $3000\pm100$    & -              & -             & M$5.0\pm0.5$ & 0.8675          & 0.00  & 17  \\
  19f-4-05194a       & $0.531\pm0.016$   & $0.651\pm0.007$   & $4400\pm100$    & -              & -             & -            & 0.5895          & 0.00  & 17  \\
  19f-4-05194b       & $0.385\pm0.011$   & $0.425\pm0.006$   & $3500\pm100$    & -              & -             & M$2.5\pm0.5$ & 0.5895          & 0.00  & 17  \\
  NGTS\,J0522-2507a  & $0.1739\pm0.0015$ & $0.2045\pm0.0058$ & $2995\pm100$    & -              & -             & M$5.0\pm1.0$ & 1.7477          & 0.00  & 18    \\
  NGTS\,J0522-2507b  & $0.1742\pm0.0019$ & $0.2168\pm0.0048$ & $2997\pm100$    & -              & -             & M$5.0\pm1.0$ & 1.7477          & 0.00  & 18    \\
  Kepler-16B         & $0.2026\pm0.0007$ & $0.2262\pm0.0006$ & -               & $-0.30\pm0.20$ & -             & -            & 41.0792         & 0.16  & 19    \\
  Kepler-38B         & $0.249\pm0.010$   & $0.2724\pm0.0050$ & -               & -              & -             & -            & 18.7953         & 0.10  & 20    \\
  Kepler-47B         & $0.362\pm0.013$   & $0.3506\pm0.0063$ & $3357\pm100$    & -              & -             & -            & 7.4484          & 0.02  & 21    \\
  PH1B               & $0.378\pm0.023$   & $0.408\pm0.024$   & -               & -              & -             & -            & 20.0003         & 0.22  & 22    \\
  Kepler-453B        & $0.195\pm0.002$   & $0.2150\pm0.0014$ & $3226\pm100$    & -              & -             & -            & 27.3220         & 0.05  & 23    \\
  \multicolumn{2}{l}{\bf Secondaries in eclipsing binaries:}\\
  OGLE-TR\,5b        & $0.271\pm0.035$   & $0.263\pm0.012$   & -               & -              & -             & -            & 0.8083          & 0.00  & 11  \\ 
  OGLE-TR\,6b        & $0.359\pm0.025$   & $0.393\pm0.018$   & -               & -              & -             & -            & 4.5488          & 0.00  & 11  \\
  OGLE-TR\,7b        & $0.281\pm0.029$   & $0.282\pm0.013$   & -               & -              & -             & -            & 2.7182          & 0.00  & 11  \\
  OGLE-TR\,18b       & $0.387\pm0.049$   & $0.39\pm0.04$     & -               & -              & -             & -            & 2.2280          & 0.00  & 11  \\
  OGLE-TR\,34b       & $0.509\pm0.038$   & $0.435\pm0.033$   & -               & $+0.32\pm0.31$ & -             & -            & 8.5763          & 0.00  & 11  \\
  OGLE-TR\,78b       & $0.243\pm0.015$   & $0.240\pm0.013$   & -               & -              & -             & -            & 5.3187          & 0.12  & 11  \\
  OGLE-TR\,106b      & $0.116\pm0.021$   & $0.181\pm0.013$   & -               & -              & -             & -            & 2.5359          & 0.00  & 11  \\
  OGLE-TR\,120b      & $0.47\pm0.04$     & $0.42\pm0.02$     & -               & -              & -             & -            & 9.1662          & 0.36  & 11  \\
  OGLE-TR\,122b      & $0.092\pm0.009$   & $0.120\pm0.019$   & -               & $+0.15\pm0.36$ & -             & -            & 7.2695          & 0.23  & 11  \\
  OGLE-TR\,123b      & $0.085\pm0.011$   & $0.133\pm0.009$   & -               & -              & -             & -            & 1.8039          & 0.00  & 24  \\
  OGLE-TR\,125b      & $0.209\pm0.033$   & $0.211\pm0.027$   & -               & -              & -             & -            & 5.3039          & 0.00  & 11  \\
  HAT-TR-205-013b    & $0.124\pm0.010$   & $0.167\pm0.006$   & -               & $-0.20\pm0.50$ & -             & -            & 2.2307          & 0.01  & 25  \\
  KOI-189b           & $0.0745\pm0.0033$ & $0.1025\pm0.0024$ & -               & $-0.12\pm0.10$ & $6.9_{-3.4}^{+6.4}$ & -            & 30.360          & 0.28  & 26  \\
  KOI-686b           & $0.0987\pm0.0049$ & $0.1250\pm0.0038$ & -               & $-0.06\pm0.13$ & $6.20\pm2.80$ & -            & 52.514          & 0.56  & 26  \\
  EPIC\,202900527b   & $0.1459\pm0.0030$ & $0.1702\pm0.0046$ & -               & $+0.32\pm0.04$ & $8.49_{-1.35}^{+0.97}$ & -            & 13.009          & 0.38  & 27  \\
  EPIC\,206155547b   & $0.1612\pm0.0070$ & $0.1996\pm0.0119$ & -               & $-0.32\pm0.04$ & $10.4_{-1.0}^{+1.2}$  & -            & 24.388          & 0.36  & 27  \\
  EPIC\,206432863b   & $0.0942\pm0.0019$ & $0.0913\pm0.0048$ & -               & $+0.01\pm0.04$ & $9.16_{-0.91}^{+0.93}$ & -            & 11.990          & 0.26  & 27  \\
  SAO\,106989b       & $0.256\pm0.005$   & $0.326\pm0.012$   & $2380\pm260$    & $-0.20\pm0.10$ & $\sim$2.0     & -            & 4.3979          & 0.25  & 28  \\
  HD\,24465b         & $0.233\pm0.002$   & $0.244\pm0.001$   & $2335\pm10$     & $+0.30\pm0.15$ & $\sim$2.3     & -            & 7.1963          & 0.21  & 28  \\
  EPIC\,211682657b   & $0.599\pm0.017$   & $0.566\pm0.005$   & $4329\pm50$     & $-0.10\pm0.15$ & $\sim$1.4     & -            & 3.1420          & 0.01  & 28  \\
  HD\,205403b        & $0.406\pm0.005$   & $0.444\pm0.014$   & $4651\pm120$    & $-0.10\pm0.15$ & $\sim$1.2     & -            & 2.4449          & 0.00  & 28  \\
  KIC\,1571511b      & $0.1414\pm0.0004$ & $0.1783\pm0.0006$ & $4100\pm50$     & -              & -             & -            & 14.0225         & 0.33  & 29  \\
  1SWASP\,J0113+3149 & $0.186\pm0.010$   & $0.209\pm0.011$   & $3922\pm42$     & $-0.40\pm0.04$ & $9.5\pm1.0$   & -            & 14.2769         & 0.31  & 30  \\
  UCAC4 714-021661b  & $0.188\pm0.014$   & $0.234\pm0.009$   & -               & $-0.05\pm0.17$ & -             & -            & 1.3512          & 0.07  & 31  \\
  T-Cyg1-12664b      & $0.376\pm0.017$   & $0.3475\pm0.0081$ & $3460\pm210$    & -              & -             & M$3.0$       & 4.1288          & 0.04  & 32  \\
  EBLM\,J0555-57Ab   & $0.0813_{-0.0037}^{+0.0038}$ & $0.084_{-0.004}^{+0.014}$ & -   & $-0.24\pm0.16$ & -             & -            & 7.7577          & 0.09  & 33  \\
  TYC 7760-484-1b    & $0.091\pm0.002$   & $0.117\pm0.006$   & -               & $-0.21\pm0.07$ & $6-12$        & -            & 6.7600          & 0.06  & 34  \\
  EPIC\,219654213    & $0.187_{-0.013}^{+0.012}$ & $0.200_{-0.008}^{+0.007}$ & -      & $-0.08\pm0.09$ & $4.1\pm1.1$   & M$5.0$       & 5.4420          & 0.01  & 35  \\
  \multicolumn{2}{l}{\bf Single stars:}\\
  GJ\,15A            & $0.405\pm0.041$   & $0.3863\pm0.0021$ & $3602\pm13$     & $-0.30\pm0.07$ & -             & M$1.5\pm0.5$ & -               & -     & 36  \\
  GJ\,205            & $0.637\pm0.064$   & $0.5735\pm0.0044$ & $3850\pm22$     & $+0.49\pm0.07$ & -             & M$0.0\pm1.0$ & -               & -     & 36  \\
  GJ\,380            & $0.711\pm0.071$   & $0.6398\pm0.0046$ & $4176\pm19$     & $+0.24\pm0.07$ & -             & -            & -               & -     & 36  \\
  GJ\,526            & $0.490\pm0.049$   & $0.4840\pm0.0084$ & $3646\pm34$     & $-0.31\pm0.07$ & -             & M$1.5\pm0.5$ & -               & -     & 36  \\
  GJ\,687            & $0.403\pm0.040$   & $0.4183\pm0.0070$ & $3457\pm35$     & $-0.05\pm0.07$ & -             & M$3.0\pm0.5$ & -               & -     & 36  \\
  GJ\,880            & $0.572\pm0.057$   & $0.5477\pm0.0048$ & $3731\pm16$     & $+0.21\pm0.07$ & -             & M$1.5\pm0.5$ & -               & -     & 36  \\
  GJ\,887            & $0.494\pm0.049$   & $0.4712\pm0.0086$ & $3695\pm35$     & $-0.06\pm0.07$ & -             & M$2.0\pm1.0$ & -               & -     & 36  \\
  GJ\,699            & $0.159\pm0.016$   & $0.1869\pm0.0012$ & $3238\pm11$     & $-0.40\pm0.07$ & -             & M$4.0$       & -               & -     & 36  \\
  GJ\,411            & $0.392\pm0.039$   & $0.3924\pm0.0033$ & $3532\pm17$     & $-0.38\pm0.07$ & -             & M$2.0$       & -               & -     & 36  \\
  GJ\,105A           & $0.767\pm0.124$   & $0.7949\pm0.0062$ & $4704\pm21$     & $-0.28\pm0.07$ & -             & -            & -               & -     & 36  \\
  GJ\,338A           & $0.630\pm0.063$   & $0.5773\pm0.0131$ & $3953\pm41$     & $-0.01\pm0.07$ & -             & -            & -               & -     & 36  \\
  GJ\,338B           & $0.617\pm0.062$   & $0.5673\pm0.0137$ & $3926\pm37$     & $-0.04\pm0.07$ & -             & -            & -               & -     & 36  \\
  GJ\,412A           & $0.390\pm0.039$   & $0.3982\pm0.0091$ & $3537\pm41$     & $-0.37\pm0.07$ & -             & M$2.0$       & -               & -     & 36  \\
  GJ\,436            & $0.447\pm0.045$   & $0.4546\pm0.0182$ & $3520\pm66$     & $+0.01\pm0.07$ & -             & M$2.5\pm0.5$ & -               & -     & 36  \\
  \hline
  \multicolumn{10}{r}{\it continues on the next page...}\\
\end{tabular}
\end{table*}

\begin{table*}
 \centering
  \tabcolsep=0.3cm
  \begin{tabular}{@{}lccccccccc@{}}
  \hline
  Name               & Mass              & Radius            &  $T_\mathrm{eff}$ & [Fe/H]         & Age           & Sp type     &  P$_\mathrm{orb}$ & $e$    & Ref \\
                     & (\MSUN)           & (\RSUN)           & (K)             & (dex)          & (Gyr)         &              & (days)          &        &     \\
  \hline
  GJ\,570A           & $0.740\pm0.119$   & $0.7390\pm0.0190$ & $4588\pm58$     & $-0.06\pm0.07$ & -             & -            & -               & -     & 36  \\
  GJ\,581            & $0.308\pm0.031$   & $0.2990\pm0.0100$ & $3487\pm62$     & $-0.15\pm0.07$ & -             & M$3.0$       & -               & -     & 36  \\
  GJ\,702B           & $0.749\pm0.075$   & $0.6697\pm0.0089$ & $4475\pm33$     & $+0.01\pm0.07$ & -             & -            & -               & -     & 36  \\
  GJ\,752A           & $0.330\pm0.033$   & $0.3561\pm0.0039$ & $3417\pm17$     & $-0.23\pm0.07$ & -             & M$3.0\pm0.5$ & -               & -     & 36  \\
  GJ\,809            & $0.573\pm0.057$   & $0.5472\pm0.0067$ & $3744\pm24$     & $-0.06\pm0.07$ & -             & M$1.0\pm1.0$ & -               & -     & 36  \\
  GJ\,820A           & $0.727\pm0.073$   & $0.6611\pm0.0048$ & $4399\pm16$     & $-0.27\pm0.07$ & -             & -            & -               & -     & 36  \\
  GJ\,820B           & $0.656\pm0.066$   & $0.6010\pm0.0072$ & $4025\pm24$     & $-0.22\pm0.07$ & -             & -            & -               & -     & 36  \\
  GJ\,892            & $0.771\pm0.124$   & $0.7784\pm0.0053$ & $4773\pm20$     & $-0.23\pm0.07$ & -             & -            & -               & -     & 36  \\
  GJ\,514            & $0.526\pm0.053$   & $0.611\pm0.043$   & $3377\pm100$    & $-0.24\pm0.20$ & -             & M$1.0\pm1.0$ & -               & -     & 11  \\
  GJ\,191            & $0.281\pm0.014$   & $0.291\pm0.025$   & $3570\pm156$    & $-0.90\pm0.20$ & -             & M$3.5\pm1.0$ & -               & -     & 11  \\
  GJ\,551            & $0.123\pm0.006$   & $0.145\pm0.011$   & $3042\pm117$    & -              & -             & M$5.5\pm0.5$ & -               & -     & 11  \\
  \hline
\end{tabular}
\end{table*}

\bsp
\label{lastpage}
\end{document}